\documentclass[final,journal,letterpaper]{IEEEtran}

 \usepackage[hidelinks]{hyperref}
 \usepackage{varioref} 
 \usepackage{wrapfig} 
 \usepackage{threeparttable} 
 \usepackage{dcolumn} 
  \newcolumntype{d}{D{.}{.}{-1}}
 \usepackage{nomencl} 
  \makeglossary
 \usepackage{subfigmat} 
 \usepackage{fancyvrb} 
  \fvset{fontsize=\footnotesize,xleftmargin=2em}
 \usepackage{lettrine} 
 \usepackage{multicol}
\usepackage{lipsum}
\usepackage{mathtools}
\usepackage{cuted}
 \usepackage{epsfig}
 \usepackage{epstopdf}
 \usepackage{subfigure}
 \usepackage{latexsym}
 \usepackage{amsmath}
 \usepackage{amsfonts}
 \usepackage{amssymb}
 \usepackage{mathrsfs}
 \usepackage{cases}
 \usepackage{array}
 \usepackage{color}
 \usepackage{soul}
 \usepackage{array}
 \usepackage{verbatim}
 \setulcolor{red}
 \sethlcolor{yellow}

 \usepackage{url}
 \usepackage[noadjust]{cite}
 \usepackage{graphicx}
 \usepackage{psfrag}
 \usepackage{color}
 \usepackage[dvipsnames]{xcolor}
   \usepackage{multirow}
 \usepackage{ragged2e}
 \usepackage{graphicx}  
 \usepackage{breqn}
 \usepackage{algorithm} 
\usepackage{algpseudocode} 
 
\newcommand{\specialcell}[2][c]{%
  \begin{tabular}[#1]{@{}c@{}}#2\end{tabular}}

\newtheorem{remark}{Remark}

\usepackage{tikz}
\usepackage{soul}
\newcommand\copyrighttext{%
  \footnotesize This work has been submitted to the IEEE for possible publication. Copyright may be transferred without notice, after which this version may no longer be accessible.}
\newcommand\copyrightnotice{%
\begin{tikzpicture}[remember picture,overlay]
\node[anchor=south,yshift=10pt] at (current page.south) {\fbox{\parbox{\dimexpr\textwidth-\fboxsep-\fboxrule\relax}{\copyrighttext}}};
\end{tikzpicture}%
}

\begin{document}

\title{Distributed-MPC with Data-Driven Estimation of Bus Admittance Matrix in Voltage Control}

\author{Ramij R. Hossain,~\IEEEmembership{Student Member,~IEEE,}
        Ratnesh Kumar,~\IEEEmembership{Fellow,~IEEE}
\thanks{The work was supported in part by the National Science Foundation under the grants, CSSI-2004766 and PFI-2141084.
\newline \indent R. R. Hossain and R. Kumar are with the Department of Electrical and
Computer Engineering, Iowa State University, Ames, IA 50011, USA (e-mail:
rhossain@iastate.edu, rkumar@iastate.edu).}
\vspace*{-.3in}
}

\maketitle
\copyrightnotice
\begin{abstract}
This article presents a {\em distributed} model-predictive control (MPC) design for real-time voltage control in power systems, including an online method to estimate the bus admittance matrix $\mathbf{Y}$ to let it be time-varying and unknown a priori. The prevalent control designs are either (a) {\em centralized}, providing optimal solutions but less scalable and susceptible to single-point failures/attacks, or (b) {\em decentralized or localized}, having increased scalability and attack resilience but are suboptimal. The proposed {\em distributed} solution offers the attractive features of both methodologies, where neighboring nodes share state information to attain a globally optimal solution. In addition, the presented framework provides a data-driven estimation of Y to circumvent the challenging issue of acquiring accurate knowledge of the line impedance (required to form Y). We first introduce the centralized version of the predictive voltage control problem and then transfer it to a distributed version. The distributed version is solved via the alternating direction method of multipliers (ADMM), using only local measurements and communication leveraging the graph structure of the power system. The proposed framework is resilient to prediction uncertainty, modeling error, and communication link failure, and the in-built redundancy within the proposed framework supports anomaly detection in cyberattacks. We validate the proposed methodology for IEEE-30 bus, IEEE-57 bus transmission systems, and IEEE-123 bus distribution systems.
\end{abstract}

\section{Introduction} \label{1}
\subsection{Motivation and Related works}
\IEEEPARstart{I}{n} order to achieve carbon-free-energy (or net-zero-energy) by 2050 \cite{netzero-DOE}, the energy regulators all over the world are pushing independent system operators (ISOs) and power utilities to adopt energy generation from renewable sources such as solar and wind, making the energy sources unpredictable and simultaneously distributed throughout the grid. With this increased penetration of variable, unpredictable, and intermittent electricity generation coupled with load uncertainties owing to dynamic loads, plug-in electric vehicles, voltage issues are becoming a predominant problem in power system operation. Traditionally, voltage control problems are solved using (a) centralized model-based algorithms \cite{valverde2013model,246}, and (b) local decentralized algorithms \cite{250}. While centralized computations are suitable for the optimal solution, but (a) are prone to single point failure, and cyber-attacks, (b) lack scalability with an increase in the number of decision variables, and (c) need robust communication infrastructure and computation resources capable of handling high volume of data. In contrast, local decentralized control is scalable and not susceptible to single-point failures/attacks, but it provides non-optimal solutions. Moreover, at times the solution provided by the local droop-control scheme is even infeasible for the voltage control problem \cite{zhu2015fast,li2014real,patari2021}. Distributed solution alternatives involving local computations based on shared information among neighbors offer optimality, scalability, as well as resiliency. This has resulted in recent studies \cite{droflersurvey,antoniadou2017distributed,niloy,sun} showing that distributed optimization and controls are emerging as potential alternatives in various sectors of power system operations.

\paragraph{Distributed Approaches and their limitations} In general, two different classes of distributed algorithms can be found in the literature for voltage control \cite{niloy}. These are: (a) static optimization algorithms and (b) dynamic optimization algorithms (also known as ``offline" and ``online" algorithms, respectively). Static (or offline) approaches \cite{Baosen,zheng2015fully,dall,Robbins2016OptimalRP,vsulc2014optimal,pac}, according to \cite{patari2021}, decompose the centralized problem into several sub-problems, and agents go through several communication rounds among neighboring agents to reach consensus over common variables using various dual-decomposition techniques, for instance, the dual-ascent method and alternating direction method of multipliers (ADMM). These approaches are open-loop in nature and typically perform optimal power flow (OPF)-based analysis with voltage constraints. It should be noted that static approaches do not use real-time measurements, such as voltage magnitudes. Among the static approaches, \cite{Baosen} utilized a semi-definite programming (SDP)-relaxed OPF formulation in conjunction with the dual-ascent method to design the distributed controllers for voltage regulation. Another SDP relaxed distributed formulation with ADMM-based decomposition can be found in \cite{dall} for optimizing active and reactive power set-points of PV inverters. In \cite{zheng2015fully}, distributed voltage control is achieved by combining ADMM with second-order cone programming (SOCP)-relaxed OPF. \cite{Robbins2016OptimalRP} presented an ADMM-based distributed reactive power compensation problem. This work utilized convex-relaxed OPF, where the nonlinear terms were held constant and updated periodically based on the desired operating point. \cite{vsulc2014optimal} presented a distributed voltage control problem minimizing line losses. This method utilized ADMM for Linearized DistFlow (LinDistFlow) model of radial-distribution network.
In \cite{pac} a distributed algorithm named proximal atomic coordination (PAC) is presented for OPF problems in a single phase distribution network. 

In contrast, dynamic (or online) approaches \cite{bolognani2014distributed,liu2017distributed,nali,magnusson2020distributed,fan2016distributed,li2019distributed, patari2022} are predominantly closed-loop, which utilize the local real-time (or current) measurements at time instant $t$ and compute the control actions for next instant $t+1$, while communicating among their neighbours. Note that to decide on the control inputs at $t+1$, the controllers can go through multiple rounds of communications to reach convergent decisions \cite{patari2022}. In this category of works, \cite{bolognani2014distributed} presents a distributed feedback algorithm to find the optimal reactive power settings for micro-generators with voltage constraints. The algorithm presented uses (a) the Lagrangian-dual method using a generalization of DC power flow model for a single-phase distribution system, and (b) relaxation of hard constraints on reactive power. In \cite{liu2017distributed}, an ADMM-based distributed method is presented for voltage control in a multi-phase distribution system with: (a) linearized sensitivity-based (linear q-v) model to avoid the computational challenges, (b) feedback-based online correction to tackle the approximations introduced, and (c) a relaxation in hard voltage constraint using a soft penalty. The authors in \cite{nali} presented a primal-dual-based distributed algorithm exploiting network sparsity with the Linearized DistFlow (LinDistflow) model of the distribution system. A similar approach can be found in \cite{magnusson2020distributed} which used asynchronous dual decomposition for faster convergence in the presence of asynchronous and delayed communication. An extension of \cite{nali} and \cite{magnusson2020distributed} can be found for multi-phase unbalanced distribution network respectively in \cite{patari2021} and \cite{patari2022}. A distributed reactive power-sharing scheme for online micro-grid voltage control with an assumption of a small phase angle difference among neighboring buses and event-triggered communication is presented in \cite{fan2016distributed}. A dual-ascent-based online distributed voltage control algorithm can be found in \cite{li2019distributed} with active power and reactive power control of the Photovoltaic (PV) system for a radial distribution network. A distributed feedback controller using the dual-ascent method was proposed in \cite{todescato} with a combined objective of maximizing the distance of voltage collapse and voltage regulation. This framework relies on the decoupling assumption (Assumption-2 of \cite{todescato}), stating that the voltage angle differences are very small (approximately 0) and remain constant. This framework, however, is not limited to the assumption of radial (acyclic) network topology, and can be used for high voltage transmission networks.

With the above summary of the state-of-art, the following issues in the existing implementations remain:
\begin{enumerate}
    \item Static (or offline) approaches utilizing SDP/SOCP-based relaxation can account for nonlinear power flow, but their (a) relaxations of the nonconvex OPF problem do not guarantee the feasibility of the obtained solution, requiring those to be verified separately \cite{niloy}, (b) methods are not computationally efficient thereby are not amenable to real-time implementation \cite{patari2022}. Our framework is more general by being dynamic, and is also free from the above limitations.
    \item Dynamic (or online) approaches of distributed voltage control are mostly focused on radial (acyclic) distribution system with Linearized DistFlow (LinDistFlow) model of power system except \cite{todescato}, which is tested for transmission systems with (a) linearized power flow model, and (b) decoupling assumption that may be restrictive in practical application. Our proposed framework is applicable to general power networks with no model assumptions/restrictions.
    \item Existing dynamic (or online) methods use the current measurements, but do not incorporate the future predictions of generation and/or load profile in control computation. This becomes limiting under uncertain and intermittent renewable generations and load variations. To tackle this, we incorporated a predictive framework.
    \item Finally, the existing methods rely on the system information, namely the bus admittance matrix $\mathbf{Y}$. But network updates/reconfiguration due to faults alter $\mathbf{Y}$ even while the system is in operation \cite{latif}. Our proposed framework estimates system parameters in runtime.
\end{enumerate}
\paragraph{Predictive Framework} Driven by availability of data logs, the importance of predictive control is growing, and is supported by the recent advances in short term forecasting \cite{wang2019review,liu2020comparative,khotanzad2002neuro,quan}. To this end, we leverage MPC, which computes the control variables iteratively at each control instant, optimizing the predicted future behavior of the underlying system, to integrate future prediction of renewable generation/load profiles with the control computation. 
MPC-based voltage control problems are mostly studied in emergency control perspective (following severe system disturbances) both in (a) centralized \cite{cutsem_review,jin2009model} and (b) distributed \cite{moradzadeh2012voltage,rrh,gobel4017547consensus} protocol. Besides, distributed-MPC-based secondary voltage control is presented in \cite{lou2016distributed}. The authors in \cite{wang2022mpc} have presented an MPC-based cluster-wise decentralized algorithm to address voltage problems utilizing electric vehicle (EV) and photovoltaic (PV) coordination. A sensitivity-based distributed MPC with leader-follower consensus protocol is presented in \cite{guo2019distributed}, where sensitivities are computed offline using a centralized protocol, as opposed to desired online using local communication.

\paragraph{Data-driven Approaches} At the same time, an interest in data-driven estimation in control design is growing in power system applications \cite{nowak1,ye2021data,hong2021data,8586174}. The authors in \cite{nowak1,ye2021data} focused on measured-data based sensitivity model computation for centralized voltage control. In \cite{nowak}, an extension of \cite{nowak1} finds area-wise linear sensitivity models of a distribution network, and further utilizes those to determine power set-points of distributed energy resources (DERs), adopting ADMM-based distributed methods. In general, a sensitivity formulation ought to employ system-wide communication, but the proposed per-area sensitivity models in \cite{nowak} take into account only one-hop communications over adjacent areas, and moreover, employ only a hypothesized linear sensitivity model. Compared to these approaches, our predictive framework can be implemented without estimating any sensitivity model, and also it does not require the knowledge of the admittance matrix $\mathbf{Y}$, rather estimates it using the measured data for which one-hop communications suffice.
Other data-driven approaches such as those involving multi-agent deep reinforcement learning (DRL) are also being developed \cite{9076841,wang2021multi}, but the learning based controllers are yet to be proven viable in a real-world application due to a lack of safety guarantee \cite{chen2021reinforcement}. In contrast, predictive control as adopted here, is well-accepted and a proven technique in many applications. 


\subsection{Our Approach and Contributions}
This article presents a novel data-driven distributed predictive voltage control framework (D3PVC) that integrates three methods:  data-driven estimation, distributed optimization, and predictive control. Instead of the previously used linearized version of the AC power flow model, which either assumes angle differences between connected buses to be small (hence, $\sin(\theta_i - \theta_k) \approx \theta_i - \theta_k$ for any two adjacent buses $i$ and $k$) or assumes relaxed DistFlow/LinDistFlow model without any constraint to ensure consistency in the voltage angles, our approach (a) takes into account the real-time measurement at current time instant $t$, (b) linearizes the nonlinear AC power flow equations with respect to the measured operating point, and (c) incorporates the predicted values of renewable generations and loads based on short-term forecasting for time instant $t+1$, (d) formulates the voltage regulation problem for incremental changes in the system (voltage) and control variables (reactive power provided by DERs) considering hard constraints in voltage and reactive power, (e) decomposes and solves the optimization problem in a distributed manner adopting ADMM and relying on communications among local neighbors, and (f) finally, allows unknown line resistances and reactances: We propose a method for data-driven estimation of the bus admittance matrix $\mathbf{Y}$ using the real-time SCADA/PMU data that is also performed distributively involving local communication among neighboring agents. 
In summary, our main contributions are as follows:
\begin{enumerate}
    \item We developed a dynamic (or online) distributed voltage control framework that is general: Not limited to the assumption of the radial network, and can be applicable for transmission and distribution systems for both meshed and radial networks; allows time-varying generation/load/network-parameters/-connectivity.
    \item The proposed distributed method is (a) predictive  by integrating generation and load forecasting in control computation, and (b) data-driven by performing data-informed estimation of system parameters that are not required to be known a priori.
   \item It incorporates a detailed sensitivity analysis for parameters in the optimization algorithms to balance accuracy and speed. 
   \item Owing to successive linearization, the resulting optimization problem becomes convex. Therefore, the ADMM-based distributed solution guarantees asymptotic convergence to the globally optimal solution \cite{melanie,boyd2011distributed}. 
   \item The proposed method is robust to (a) prediction errors in generation and load forecasting, (b) modeling errors owing to data-driven estimation, and (c) measurement noises.
   \item The proposed method is resilient to communication failures, and network attacks.
   \item We also extended the proposed framework from bus-wise distributed computation to area-wise distributed computation.
\end{enumerate}

The proposed D3PVC methodology is tested for IEEE 30 Bus, 57 Bus transmission network, and IEEE 123 Bus distribution network to (a) achieve the desired voltage performance under data-driven estimation of $\mathbf{Y}$, (b) satisfy the network constraints, and (c) maintain the reactive power constraints throughout, while only employing local exchange of information.

\section{Power System Model and Problem Formulation}\label{2}
A power system connectivity of buses via the lines can be represented by an undirected network graph $\mathcal{G}=\{\mathbb{N},\mathbb{E}\}$, where the set of buses includes $\mathbb{N} = \{0,\cdots,N\}$ and $\mathbb{E}$ represents the set of lines connecting the buses, with line $l_{ik}\in\mathbb{E}$ connecting buses $i,k\in \mathbb{N}$. The magnitude and angle of voltage at bus $i$ are denoted $V_i$ and $\theta_i$ respectively. We let bus 0 to be the slack/reference bus, so by convention, $V_0 = 1$ p.u. and $\theta_0 = 0^{\circ}$.  The symmetric bus admittance matrix $\mathbf{Y} =[Y_{ik}] \in \mathbf{C^{|\mathbb{N}|\times |\mathbb{N}|}}$ of the given network is defined as:
\begin{equation}\label{adm_mat}
Y_{ik} =
\begin{cases}
      y_{i} + \sum_{l=0,l\neq i}^{N} y_{il}, & \text{if\;\;$i=k$}\\
      -y_{ik} & \text{otherwise}
\end{cases},
\end{equation} 
where  $y_i$ denotes the (complex) admittance to the ground at bus $i$, $y_{ik}$ is the admittance of line $l_{ik} \in \mathbb{E}$ (implying $y_{ik} = y_{ki} \neq 0$ if $l_{ik} \in \mathbb{E}$ and otherwise $y_{ik} = y_{ki} = 0$). We write the complex admittance as, $Y_{ik} =  G_{ik} + jB_{ik}$, with $G_{ik} :=$ real part (conductance) and $B_{ik} :=$ imaginary part (susceptance).

The active and reactive power generations at any bus $i$ are denoted by $P^G_i, Q^G_i$, respectively, while $P^D_i, Q^D_i$ represent the active and reactive power demands, respectively. Therefore the net injections of active and reactive powers at bus $i$ are given by: $P_i^{\text{in}} = P_i^{G} - P_i^{D}$, and $Q_i^{\text{in}} = Q_i^{G} - Q_i^{D}$.
Letting $\mathbb{N}_i:=\{k:l_{ik}\in\mathbb{E}\}$ denote the neighboring buses that are one hop away from bus $i$, the power flow relation at bus $i$ is given by:
\begin{subequations}\label{pfeqn}
\begin{gather}
    P_i^{\text{in}} =\sum_{k\in\{i\}\cup \mathbb{N}_i}V_iV_k\Big[G_{ik}\cos{(\theta_i-\theta_k)} + B_{ik}\sin{(\theta_i-\theta_k)}\Big],\nonumber\\
    \hspace*{-1.1in}\equiv g^P_i(V_i,\theta_i,V_k,\theta_k\mid k\in\mathbb{N}_i),\\
    Q_i^{\text{in}} =  \sum_{k\in \{i\}\cup\mathbb{N}_i}V_iV_k\Big[G_{ik}\sin{(\theta_i-\theta_k)} - B_{ik}\cos{(\theta_i-\theta_k)}\Big]\nonumber\\
    \hspace*{-1.1in}\equiv g^Q_i(V_i,\theta_i,V_k,\theta_k\mid k\in\mathbb{N}_i).
\end{gather}
\end{subequations}

With the proliferation of renewable energy generation and the variable nature of the loads, maintaining voltage trajectories close to the desired reference value of $V_{\text{ref}} = 1.00$ p.u. is challenging, requiring a well-defined control design.  
For this, we proposed to leverage the recent developments in data-driven prediction techniques, through which an accurate prediction of up to an hour-ahead generation and load profiles commonly referred as short-term forecasting, has become possible for most utilities and system operators \cite{wang2019review,liu2020comparative,khotanzad2002neuro,quan}. The prediction error is typically within 3-5\% for the case of data collected at California ISO \cite{caiso}. Short-term forecasting (STF) of load and renewable generation has found success in day-to-day operations of power system, especially in unit commitment (UC), scheduling operations, system security and control of power systems \cite{quan}. As an example, \cite{agalgaonkar2013distribution} proposed a reactive power control strategy utilizing load and solar irradiance forecast, while \cite{trudnowski2001real} designed a real time automatic generation control (AGC) strategy based on short term load prediction.

In line with the above works, even in our work presented here, we employ the predicted generation and load profiles available at the utilities along with the latest measurements of the system variables: At any control time instant $t$, the bus voltages and angles $[V^t_i,\theta^t_i]$ as well as the active and reactive power injections $[P_i^{\text{in},t},Q_i^{\text{in},t}]$ are measured and hence known for all $i\in \mathbb{N}$, while additionally, the predicted generations $[\widehat{P}^{G,t+1}_i,\widehat{Q}^{G,t+1}_i]$ and predicted demand profiles $[\widehat{P}^{D,t+1}_i,\widehat{Q}^{D,t+1}_i]$, and thereby the predicted injections $[\widehat{P}_i^{\text{in},t+1},\widehat{Q}_i^{\text{in},t+1}]$ for the next control time instant $t+1$ are also available from the forecast data.


Under the knowledge of the above data at control instant $t$, we next present a predictive control formulation to compute an optimal reactive power correction at the designated buses for time instant $t+1$:
For predicted injections of active and reactive powers $[\widehat{P}_i^{\text{in},t+1},\widehat{Q}_i^{\text{in},t+1}]$ and a reactive compensation of $u_i^{t+1}$ (a control variable to be decided through optimization), the power flow equations (\ref{pfeqn}) at bus $i$ for time $t+1$ become: 
\begin{subequations}\label{pfneighnext}
\begin{gather}
\!\!\!\!\!\!\!\!\!\!\!\!g^P_i(\widehat{V}^{t+1}_i,\widehat{\theta}^{t+1}_i,\widehat{V}^{t+1}_k,\widehat{\theta}^{t+1}_k\mid k\in\mathbb{N}_i)=\widehat{P}_i^{\text{in},t+1},\\ 
\!\!\!\!g^Q_i(\widehat{V}^{t+1}_i,\widehat{\theta}^{t+1}_i,\widehat{V}^{t+1}_k,\widehat{\theta}^{t+1}_k\mid k\in\mathbb{N}_i)= \widehat{Q}_i^{\text{in},t+1} + u_i^{t+1},
\end{gather}
\end{subequations}
where 
$[\widehat{V}^{t+1}_k= V_k^{t} + \Delta V_k^{t+1}, \widehat{\theta}^{t+1}_k= \theta_k^{t} + \Delta \theta_k^{t+1}],k \in\{i\}\cup \mathbb{N}_i$) are {\em predicted} values for bus $i$ and its neighboring buses in $\mathbb{N}_i$ at $t+1$, and satisfy Equations~(\ref{centmpc2})-(\ref{centmpc3}).

Expressing the {\em predicted} changes in injected active and reactive powers as: 
\begin{equation*}
\Delta P_i^{\text{in},t+1}=\widehat{P}_i^{\text{in},t+1} - P_i^{\text{in},t};
\Delta Q_i^{\text{in},t+1}=\widehat{Q}_i^{\text{in},t+1} - Q_i^{\text{in},t},
\end{equation*}
and applying \emph{Taylor series approximation} on (\ref{pfneighnext}), we obtain:
\small
\begin{subequations}\label{pfneightaylor}
\begin{multline}\label{centmpc2}
\Big(\frac{\partial g^P_i}{\partial V_i}\Big)^t\Delta V_i^{t+1} + \Big(\frac{\partial g^P_i}{\partial \theta_i}\Big)^t \Delta \theta_i^{t+1} + \sum_{k \in \mathbb{N}_i} \Big [\Big(\frac{\partial g^P_i}{\partial V_k}\Big)^t\Delta V_k^{t+1} + \\\Big(\frac{\partial g^P_i}{\partial \theta_k}\Big)^t\Delta \theta_k^{t+1}\Big] = \Delta P^{\text{in},t+1}_i,
\end{multline}
\begin{multline}\label{centmpc3}
\Big(\frac{\partial g^Q_i}{\partial V_i}\Big)^t\Delta V_i^{t+1} + \Big(\frac{\partial g^Q_i}{\partial \theta_i}\Big)^t \Delta \theta_i^{t+1} + \sum_{k \in \mathbb{N}_i} \Big [\Big(\frac{\partial g^Q_i}{\partial V_k}\Big)^t\Delta V_k^{t+1} + \\\Big(\frac{\partial g^Q_i}{\partial \theta_k}\Big)^t\Delta \theta_k^{t+1}\Big] = \Delta Q^{\text{in},t+1}_i + u_i^{t+1},
\end{multline}
\end{subequations}
\normalsize
in which $u_i^{t+1}$ 
(and hence also $[\Delta V_i^{t+1},\Delta \theta_i^{t+1}]$) is determined through the following proposed \emph{Centralized Predictive Voltage Control (C-PVC)}:
\vspace{-0.05 in}

\small
\begin{subequations}\label{centmpc}
\begin{gather}\label{centmpc1}
\min_{\Delta{V}^{t+1},\Delta{\theta}^{t+1},u^{t+1}}\sum_{i=1}^{N}\Big[{{\lvert\lvert V_i^t + \Delta{V}_i^{t+1} - V_{\text{ref}} \rvert \rvert}_2} + {\lvert\lvert w_iu_i^{t+1} \rvert \rvert}_2\Big],\\
\shortintertext{subject to: $\forall i\in\mathbb{N},$}\nonumber
\text{System Constraints:} \mbox{ (\ref{centmpc2})-(\ref{centmpc3})}\nonumber,\\
u_{i,\text{min}} \leq u^{t+1}_i \leq u_{i,\text{max}},\label{centmpc4} \\
V_{i,\text{min}} \leq V_i^{t} + \Delta V_i^{t+1} \leq V_{i,\text{max}},\label{centmpc5} \\
\Delta{\theta}_{i,\text{min}} \leq \Delta{\theta}_i^{t+1} \leq \Delta{\theta}_{i,\text{max}}.\label{centmpc6}
\end{gather}
\end{subequations}
\normalsize
In the C-PVC formulation of (\ref{centmpc}), the optimization variables correspond to: $\Delta{V}^{t+1} := [\Delta{V}^{t+1}_1, \cdots, \Delta{V}^{t+1}_{N-1}]^T$, $\Delta{\theta}^{t+1} := [\Delta{\theta}^{t+1}_1,\cdots,\Delta{\theta}^{t+1}_{N-1}]^T$, and $u^{t+1} := [u_1^{t+1},\cdots,u_{N-1}^{t+1}]^T$, while
$[u_{i,\text{min}},u_{i,\text{max}}],\;  [V_{i,\text{min}},V_{i,\text{max}}],\;[\Delta{\theta}_{i,\text{min}}, \Delta{\theta}_{i,\text{max}}]$ represent the lower and upper bounds for changes in control inputs (reactive compensation), the voltage values, and the changes in angles of bus $i$, respectively.

The C-PVC formulation (\ref{centmpc}) is an instance of convex optimization, with a convex objective and affine constraints; hence the problem can be solved for the unique global optimum using any efficient solver, e.g., CVX, CVXPY, CPLEX.

\section{Distributed Solution of C-PVC}\label{3}
Here our goal is to solve (\ref{centmpc}) distributively and in data-driven fashion, employing (i)  distributed computation involving communication among only the neighbors, and (ii) data to estimate $\mathbf{Y}$ (and hence of $G_{ik}$ and $B_{ik}$) required to compute the partial derivative terms of (\ref{centmpc2})-(\ref{centmpc3}). 
We start by casting the C-PVC problem (\ref{centmpc}) as an instance of $N$-agent distributed optimization, utilizing the idea of distributed estimation and consensus as in \cite{samar2007distributed,boyd2011distributed}. 
(Here we consider the communication graph among the agents to be the same as the power system network graph $\mathcal{G}$.) Subsequently, we present an ADMM-based algorithm to solve the distributed optimization problem, and finally show how to incorporate the data-driven estimation, thereby eliminating the need of knowing the system model parameters in performing the optimization. 


Due to graph structure of the power system, the constraints (\ref{centmpc2})-(\ref{centmpc3}) of any bus $i$ depends on the change of voltages and angles of bus $i$ as well as all the neighboring buses $k$ in $\mathbb{N}_i$. This also implies that an alteration of $(\Delta{V}_i^{t+1},\Delta{\theta}_i^{t+1})$ impacts not only the model equations of bus $i$, but also the model equations of bus $k \in \mathbb{N}_i$. In contrast, any change in the $i$th control input $\{u_i^{t+1}\}$ only impacts the model equation of respective bus $i$. Keeping this in mind, we separate the optimization variables in (\ref{centmpc}) into two categories: a) public variables: $\Delta{V}^{t+1},\Delta{\theta}^{t+1}$ and b) private variables: $u^{t+1}$. 

To be able to reformulate (\ref{centmpc}) as a distributed optimization, we next introduce at each bus a local replica copy of each public variable, namely, at each bus $i\in\mathbb{N}$, we introduce a vector \[x_i:=\left[\underbrace{(\Delta{V}_{k}^{t+1},\Delta{\theta}_{k}^{t+1})}_{:=x_{(i,k)}}\mid k\in \{i\}\cup\mathbb{N}_i\right]^T\in \mathbb{R}^{2\times(\lvert{\mathbb{N}_i}\rvert+1)}.\] 
In addition to $x_i$, each bus $i$ maintains a separate copy of its own local variables as the {\em true values} of the respective variables: \[z_i := \left(\Delta{V}_i^{t+1},\Delta{\theta}_i^{t+1}\right) \in \mathbb{R}^2.\] 
The true variables in $z:=[z_i|i\in\mathbb{N}]^T$ and the local replica variables in $x:=[x_i|i\in\mathbb{N}]^T$ must together satisfy the self consistency constraints expressed as follows:
\begin{subequations} \label{defE}
\begin{gather}
    \forall i\in\mathbb{N}:x_i = E_i z,\mbox{ where}\\
\forall k\!\in\!\{i\}\!\cup\!\mathbb{N}_i,l\!\in\!\mathbb{N}\!:E_{i}(k,l)\!:=\!
\begin{cases}
      \!\mathbf{I}_{2\times2}
      & \text{if\;\;$x_{(i,k)} = z_l$}\\
      \!\mathbf{0}_{2\times2} & \text{otherwise}
\end{cases}
\end{gather}
\end{subequations}
Note $E_i \in \mathbb{R}^{(2\times(\lvert{\mathbb{N}_i}\rvert+1))\times (2\times\lvert{\mathbb{N}}\rvert)}$, and if we let $E_i(k)$ denotes its $k^{\rm th}$ column $(k\in\mathbb{N})$, (\ref{defE}a) can be rewritten as (\ref{centmpcchvar2}).\\ [-8pt]

\begin{figure}[htbp]
  \centering
    \includegraphics[width=0.40\textwidth]{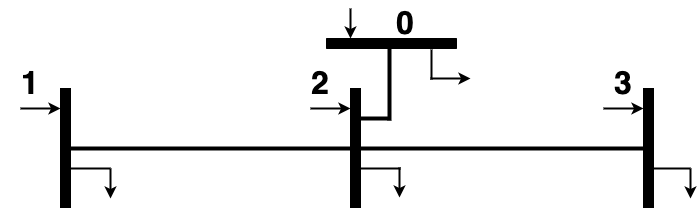}
  \caption{Illustrative Example}
  \label{f2}
 \end{figure}

\noindent\hrulefill\\
\noindent\textbf{Illustrative Example:} The distributed computation framework involving the new set of variables defined above is illustrated using a 4-Bus system shown in Fig. \ref{f2}. By the convention adopted above, bus 0 is the slack bus, implying $V_0 = 1$ p.u. and $\theta_0 = 0^{\circ}$, and these remain fixed, so that $\Delta V_0 = 0$, and $\Delta{\theta}_0 = 0$. 
\normalsize
 
The details of the local copies and true public and private variables are given next.
\begin{table}[!htbp]
\caption{Public and Private Variables}\label{tab:details1} 
\centering
\tabcolsep=0.05 cm
\begin{tabular}{|c|c|c|c|c|}  
\hline 
Bus No  & Neighbors & \specialcell{Public\\ Local Copy }& \specialcell{Public\\True}  & \specialcell{Private} \\ [0.5ex] 
\hline 
Bus-1 & $\mathbb{N}_1$ = \{2\} & $x_1 = [x_{(1,1)},x_{(1,2)}]^T$ & $z_1$ & $u_1$\\ 
\hline
Bus-2 & $\mathbb{N}_2$ =\{1,3\} & $x_2 = [x_{(2,2)},x_{(2,1)},x_{(2,3)}]^T$ & $z_2$ &$u_2$\\
\hline
Bus-3 & $\mathbb{N}_3$ =\{2\} & $x_3 = [x_{(3,3)},x_{(3,2)}]^T$ & $z_3$&$u_3$ \\
\hline
\end{tabular}
\end{table}
\small
\begin{gather*}
@\text{Bus-1:}
\begin{cases}
     x_{(1,1)} = (\Delta{V}_{1}^{t+1},\Delta{\theta}_{1}^{t+1}),\\
     x_{(1,2)} = (\Delta{V}_{2}^{t+1},\Delta{\theta}_{2}^{t+1}),
\end{cases}\\
@\text{Bus-2:}
\begin{cases}
     x_{(2,2)} = (\Delta{V}_{2}^{t+1},\Delta{\theta}_{2}^{t+1}), \\
     x_{(2,1)} = (\Delta{V}_{1}^{t+1},\Delta{\theta}_{1}^{t+1}),\\x_{(2,3)} = (\Delta{V}_{3}^{t+1},\Delta{\theta}_{3}^{t+1}),
\end{cases}\\
@\text{Bus-3:}
\begin{cases}
     x_{(3,3)} = (\Delta{V}_{3}^{t+1},\Delta{\theta}_{3}^{t+1}),\\
     x_{(3,2)} = (\Delta{V}_{2}^{t+1},\Delta{\theta}_{2}^{t+1}).
\end{cases}
\end{gather*}
\normalsize
\noindent For the self consistency of the variables we need:
\begin{equation*}
\begin{gathered}
    x_{(1,1)} = x_{(2,1)} = z_1,\\
    x_{(2,2)} = x_{(1,2)} = x_{(3,2)}= z_2,\\
    x_{(3,3)} = x_{(2,3)} = z_3.
\end{gathered}
 \end{equation*}
Hence we can define $E_i$'s as follows:

\scriptsize
\begin{equation*}
\begin{aligned}
E_1 =  \begin{bmatrix}
E_{1}(1) & E_{1}(2) & E_{1}(3)\end{bmatrix} = \begin{bmatrix}
\mathbf{I} & \mathbf{0} & \mathbf{0}\\
\mathbf{0} & \mathbf{I} & \mathbf{0} \end{bmatrix}, \\ \\
E_2 =  \begin{bmatrix}
E_{2}(1) & E_{2}(2) & E_{2}(3)\end{bmatrix} = \begin{bmatrix}
\mathbf{0} & \mathbf{I} & \mathbf{0}\\
\mathbf{I} & \mathbf{0} & \mathbf{0} \\ 
\mathbf{0} & \mathbf{0} & \mathbf{I} \end{bmatrix}, \\ \\
E_3 =  \begin{bmatrix}
E_{3}(1) & E_{3}(2) & E_{3}(3)\end{bmatrix} = \begin{bmatrix}
\mathbf{0} & \mathbf{0} & \mathbf{I}\\
\mathbf{0} & \mathbf{I} & \mathbf{0} \end{bmatrix}.
\end{aligned}
\end{equation*}
\hrulefill

\normalsize
With the above introduction of variables, the equations (\ref{centmpc2})-(\ref{centmpc3}) for bus $i$ can be represented as follows:
\begin{subequations}\label{eqnchvar1}
\begin{gather}
\sum_{k \in\{i\}\cup\mathbb{N}_i} \Big[{\nabla}_{x_{(i,k)}}{g_i^P}\Big]x^T_{(i,k)} = \Delta P^{\text{in},t+1}_i,\\
\sum_{k \in\{i\}\cup\mathbb{N}_i} \Big[{\nabla}_{x_{(i,k)}}{g_i^Q}\Big]x^T_{(i,k)} = \Delta Q^{\text{in},t+1}_i + u_i,
\end{gather}
\end{subequations}
where for $k\in\{i\}\cup\mathbb{N}_i$: 
\scriptsize
\begin{gather*} 
{\nabla}_{x_{(i,k)}}{g_i^P} = \Big[\Big(\frac{\partial g^P_i}{\partial V_k}\Big)^t \Big(\frac{\partial g^P_i}{\partial \theta_k}\Big)^t\Big]^T, 
{\nabla}_{x_{(i,k)}}{g_i^Q} = \Big[\Big(\frac{\partial g^Q_i}{\partial V_k}\Big)^t \Big(\frac{\partial g^Q_i}{\partial \theta_k}\Big)^t\Big]^T.
\end{gather*}
\normalsize
Additionally, the constraints (\ref{centmpc4})-(\ref{centmpc6}) can be transformed as follows:
\begin{equation}\label{eqnchvar2}
\begin{gathered}
u_{i,\text{min}} \leq u_i \leq u_{i,\text{max}}\;,\;x_{i,\text{min}} \leq x_i \leq x_{i,\text{max}}.
\end{gathered}
\end{equation}
The affine equality (\ref{eqnchvar1}) and the feasibility inequality (\ref{eqnchvar2}) constraints together represent a convex set $\mathbb{C}_i$ for each bus $i$, and it must hold that: $(x_i,u_i) \in \mathbb{C}_i,\forall i\in\mathbb{N}.$
Further the objective function is separable into those for individual buses, and that of bus $i\in\mathbb{N}$ can be written as: 
\begin{gather}\label{ind_cost}
   J_i(x_i,u_i):= {{\lvert\lvert V_i^t + \Delta{V}_i^{t+1} - V_{\text{ref}} \rvert \rvert}_2} + {\lvert\lvert w_iu_i^{t+1} \rvert \rvert}_2,
\end{gather}
so that: $J(x,u) = \sum_{i=1}^{N} {J_i(x_i,u_i)}$.

Following \cite{boyd2011distributed}, the distributed version of optimization (\ref{centmpc}) can then be cast as a set of local optimizations, one at each bus $i\in\mathbb{N}$:
\begin{subequations}\label{centmpcchvar}
\begin{gather}
\min_{(x_i,u_i) \in \mathbb{C}_i}\;\; {J_i(x_i,u_i)}\label{centmpcvhvar1}\\
\text{subject to,}\;\;\;\; x_i=\sum_{k \in \{i\}\cup\mathbb{N}_i} E_i(k)z_k.\label{centmpcchvar2}
\end{gather}
\end{subequations}
The augmented lagrangian for (\ref{centmpcchvar}) can be written as in (\ref{auglag}), where $\lambda$ is the lagrangian multiplier, and $\rho$ is a penalty constant: 
\begin{multline}\label{auglag}
L_{\rho}^i(x_i,u_i,z_i,\lambda_i) = {J_i(x_i,u_i)} + {\lambda_i}^T\Big(x_i - \sum_{k \in \{i\} \cup \mathbb{N}_i} E_{i}(k)z_k\Big) \\+\frac{\rho}{2}{\Big\lvert\Big\lvert x_i\;-\sum_{k \in \{i\} \cup \mathbb{N}_i} E_{i}(k)z_k \Big\rvert \Big\rvert}^2.
\end{multline}
\normalsize
Then following \cite{boyd2011distributed}, the dual problem of (\ref{centmpcchvar}) can be solved by \textbf{Algorithm} \ref{alg:admm}, which then also yields the centralized optimum of (\ref{centmpc}) as explained below in Remark~\ref{R2}.
\begin{algorithm}
  \caption{\bf Alternating Direction Method of Multiplier (ADMM)}
  \label{alg:admm}
  \begin{algorithmic}
    \State Parallely $\forall i\in\mathbb{N}$:
    \State \textbf{Intialize} $\lambda_i=z_i=0$.
    \Repeat 
    \State 1. Update $(x_i^{+},u_i^{+})=\underset{(x_i,u_i) \in \mathbb{C}_i }{\textrm{argmin}}\;\;L_{\rho}^i(x_i,u_i,z_i,\lambda_i)$
    \State 2. Communicate $x_i^{+}$ to all neighbor $k\in\mathbb{N}_i$
    \State 3. Update $z_i^{+} =  \frac{1}{\lvert{\mathbb{N}_i}\rvert+1} \sum_{k \in\{ i\}\cup \mathbb{N}_i} E_{k}(i)^T(x_{k}^{+} +\frac{1}{\rho}\lambda_{k})$
    \State 4. Communicate $z_i^{+}$ to all neighbors $k\in\mathbb{N}_i$
    \State 5. Update $\lambda_i^{+}=\lambda_i+\rho(x_i^+-\sum_{k \in\{ i\}\cup  \mathbb{N}_i} E_{i}(k)z_k$)
    \State 6. Communicate $\lambda_i^{+}$ to all neighbors $k\in\mathbb{N}_i$
     \Until convergence 
  \end{algorithmic}
\end{algorithm}

\begin{remark}\label{R2} 
Following \cite{boyd2011distributed}, whenever the local objective functions $J_i(x_i,u_i),i\in\mathbb{N}$ are {\em closed, proper, and convex} and the overall augmented Lagrangian,
\begin{equation}\label{auglag+}
    L_{\rho}(x,u,z,\lambda) = \sum_{i=1}^{N}L_{\rho}^i(x_i,u_i,z_i,\lambda_i),
\end{equation}
has a saddle point, \textbf{Algorithm} \ref{alg:admm} has the property that the residuals $\lvert\lvert x_i-E_i z\rvert\rvert$ converge asymptotically to zero for all $i\in\mathbb{N}$ and value of $J(x,u)=\sum_{i=1}^{N} {J_i(x_i,u_i)}$ converges asymptotically to the primal optimum. In our case, (a) the C-PVC problem (\ref{centmpc}) is convex, and hence its overall augmented lagrangian (\ref{auglag+}) has a saddle point and (b) the functions $J_i(x_i,u_i)$ defined in (\ref{ind_cost}) are $L_2$-norm of the optimization variables, hence they are closed, proper and convex. Therefore, the distributed solution of \textbf{Algorithm} \ref{alg:admm} converges asymptotically to the primal optimum.
\end{remark}

\section{Data-driven estimation of admittance matrix}\label{4}
The above proposed distributed optimization framework requires the computation of the required partial derivatives in (\ref{eqnchvar1}) (that appear in (\ref{centmpc2})-(\ref{centmpc3})). The partial derivatives are of the functions $g^P_i(\cdot)$ and $g^Q_i(\cdot)$ introduced in (\ref{pfeqn}), and their values depend on the entries of the admittance matrix $\mathbf{Y}$, for which from (\ref{adm_mat}), $Y_{ii}=y_{i} + \sum_{k=0,k\neq i}^{N} y_{ik}\equiv G_{ii}+jB_{ii}$. Hence we have: $G_{ii} = g_i + \sum_{k=0,k\neq i}^{N} g_{ik}= g_i -\sum_{k \in \mathbb{N}_i} G_{ik}, \mbox{ and}, B_{ii} = b_i + \sum_{k=0,k\neq i}^{N} b_{ik}= b_i -\sum_{k \in \mathbb{N}_i} B_{ik}.$ Note in general, $g_i=0$, and since the control goal is to compute the reactive compensations, $b_i$ can be absorbed in $u_i$, so that we can also treat $b_i=0$.  With this convention, we have: $G_{ii}= -\sum_{k \in \mathbb{N}_i} G_{ik}$ and $B_{ii} = -\sum_{k \in \mathbb{N}_i} B_{ik}$. Consequently, (\ref{pfeqn}) can be written as:
\begin{subequations}\label{2nddecomp}
\begin{multline}
g^P_i(V_i,\theta_i,V_k,\theta_k) = \sum_{k \in \mathbb{N}_i} \Big[\{ V_iV_k\cos{(\theta_i-\theta_k)}- V_i^2\} G_{ik} \\+ V_iV_k\sin{(\theta_i-\theta_k)}B_{ik}\Big],
\end{multline}
\vspace*{-.3in}
\begin{multline}
g^Q_i(V_i,\theta_i,V_k,\theta_k)= \sum_{k \in \mathbb{N}_i} \Big[ V_iV_k\sin{(\theta_i-\theta_k)} G_{ik} \\+\{V_i^2- V_iV_k\cos{(\theta_i-\theta_k)}\}B_{ik}\Big].
\end{multline}
\end{subequations}
We can now compute the required partial derivative terms of (\ref{pfneightaylor}), and assemble them in the matrix form (\ref{partmat1}). Clearly the values of the partial derivatives for bus $i\in\mathbb{N}$ require, a) the current measurements of bus voltages and angles of buses $\{i\}\cup\mathbb{N}_i$, and b) the knowledge of $G_{ik}$ and $B_{ik},k\in\mathbb{N}_i$, which as shown next is estimated from the same measurement data.

\small
\begin{subequations}\label{partmat1}
\begin{gather} 
\begin{bmatrix} 
\frac{\partial g^P_i}{\partial V_i}\\[5pt]
\frac{\partial g^P_i}{\partial \theta_i}\\[5pt]
\frac{\partial g^Q_i}{\partial V_i}\\[5pt]
\frac{\partial g^Q_i}{\partial \theta_i}\end{bmatrix}\!\!=\!\!\sum_{k \in \mathbb{N}_i}\!\!\begin{bmatrix}
V_k\cos{(\theta_i-\theta_k)}- 2V_i \!\!&\!\! V_k\sin{(\theta_i-\theta_k)}\\[5pt]
-V_iV_k\sin{(\theta_i-\theta_k)} \!\!&\!\! V_iV_k\cos{(\theta_i-\theta_k)}\\[5pt]
V_k\sin{(\theta_i-\theta_k)} \!\!&\!\! 2V_i- V_k\cos{(\theta_i-\theta_k)}\\[5pt]
V_iV_k\cos{(\theta_i-\theta_k)} \!\!&\!\! V_iV_k\sin{(\theta_i-\theta_k)}\end{bmatrix}\!\! \begin{bmatrix}
G_{ik}\\
B_{ik} \end{bmatrix}
\end{gather}
\vspace{-0.1in}
\begin{gather}
\begin{bmatrix}
\frac{\partial g^P_i}{\partial V_k}\\[5pt]
\frac{\partial g^P_i}{\partial \theta_k}\\[5pt]
\frac{\partial g^Q_i}{\partial V_k} \\[5pt]
\frac{\partial g^Q_i}{\partial \theta_k}\end{bmatrix}\!\!=\!\!\begin{bmatrix}
V_i\cos{(\theta_i-\theta_k)} \!\!&\!\! V_i\sin{(\theta_i-\theta_k)} \\[5pt]
V_iV_k\sin{(\theta_i-\theta_k)} \!\!&\!\! -V_iV_k\cos{(\theta_i-\theta_k)}\\[5pt] V_i\sin{(\theta_i-\theta_k)} \!\!&\!\! - V_i\cos{(\theta_i-\theta_k)}\\[5pt]
-V_iV_k\cos{(\theta_i-\theta_k)} \!\!&\!\! -V_iV_k\sin{(\theta_i-\theta_k)}\end{bmatrix}\!\! \begin{bmatrix}
G_{ik}\\
B_{ik} \end{bmatrix}\!\!.
\end{gather}
\end{subequations}
\normalsize

\begin{figure}[htbp!]
  \centering
    \includegraphics[width=0.45\textwidth]{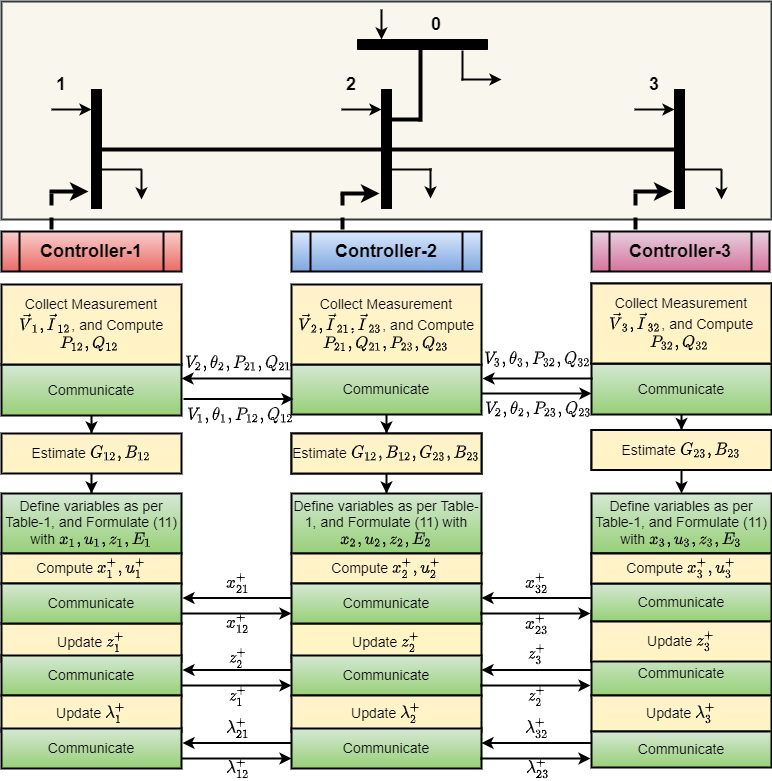}
  \caption{Flow of Information in D3PVC Framework for Illustrative Example}
  \label{ffw}
  \vspace*{-0.2in}
 \end{figure}

\noindent The complex power flowing from bus $i$ to $k$ is $S_{ik} = P_{ik} + jQ_{ik} = \vec{V_i} \vec{I}^*_{ik}$, which can be measured by measuring $\{\vec{V}_i, \vec{I}_{ik}\mid k\in\mathbb{N}_i\}$ at each bus $i\in\mathbb{N}$ from which $P_{ik}$ and $Q_{ik}$ become known. 
From (\ref{2nddecomp}) we have, $P_{ik}=\{V_iV_k\cos{(\theta_i-\theta_k)}- V_i^2\} G_{ik}+V_iV_k\sin{(\theta_i-\theta_k)}B_{ik}$ and $Q_{ik}=V_iV_k\sin{(\theta_i-\theta_k)} G_{ik}+\{V_i^2- V_iV_k\cos{(\theta_i-\theta_k)}\}B_{ik}$, that are linear functions of $G_{ik}$ and $B_{ik}$ (and so are dually $P_{ki}$ and $Q_{ki}$) and can be  written in the form of $\mathbf{A}\Theta = \mathbf{b}$, where: 
\[
    \mathbf{A} = \begin{bmatrix}
    V_iV_k\cos{(\theta_i-\theta_k)}- V_i^2 & V_iV_k\sin{(\theta_i-\theta_k)}\\
    V_iV_k\sin{(\theta_i-\theta_k)} & V_i^2- V_iV_k\cos{(\theta_i-\theta_k)}\\
    V_kV_i\cos{(\theta_k-\theta_i)}- V_k^2 & V_kV_i\sin{(\theta_k-\theta_i)} \\
    V_kV_i\sin{(\theta_k-\theta_i)} & V_k^2- V_kV_i\cos{(\theta_k-\theta_i)}
    \end{bmatrix}, 
   \]
   \[
\Theta = \begin{bmatrix}
G_{ik}\\
B_{ik} \end{bmatrix},
\;\;\text{and} \;\; 
\mathbf{b}=\begin{bmatrix}
    P_{ik}\\
    Q_{ik}\\
    P_{ki}\\
    Q_{ki}
    \end{bmatrix}.
\]
Therefore the least square estimate (LSE) \cite{watson1967linear} of $G_{ik}$ and $B_{ik}$ is 
given by the standard formula:
\begin{gather}\label{lse}
     \begin{bmatrix}
\hat G_{ik}\\
\hat B_{ik} \end{bmatrix}\equiv\hat\Theta = (\mathbf{A}^T\mathbf{A})^{-1}\mathbf{A}^T\; \mathbf{b}.
\end{gather}
The above data-driven estimation of  $\{G_{ik},B_{ik}\mid i\in\mathbb{N},k\in\mathbb{N}_i\}$ removes the need of any a priori knowledge of system parameters (the line resistances and reactances) during the optimization of (\ref{centmpcchvar}). It simply relies on the voltage and current measurements, which can be obtained from field measurement devices. Also, the direct measurements of real-time active and reactive power flow in case of transmission lines are available from SCADA measurements \cite{doeferc}, and can be utilized in conjunction with the voltage magnitude and angle measurements to solve (\ref{lse}). Thereby our proposed methodology can be executed utilizing measurement data, and hence appropriately referred to as  \textit{Data-Driven Distributed Predictive Voltage Control (D3PVC)}. The complete steps of the proposed D3PVC are captured in \textbf{Algorithm} \ref{alg:overview}. Fig.~\ref{ffw} depicts the D3PVC framework, including the underlying information flow for the illustrative example system of Fig.~\ref{f2}. 
\begin{algorithm}
  \caption{\bf Data-Driven Distributed Predictive Voltage Control (D3PVC)}
  \label{alg:overview}
  \begin{algorithmic}
    \State Parallely $\forall i\in\mathbb{N}$:
    \For {each control instant $t \in [0,T]$}
    \State 1. Obtain latest measurement data $\{\vec{V}_i, \vec{I}_{ik}\mid k\in\mathbb{N}_i\}$, and compute $P_{ik},Q_{ik}$ (recall, $\vec{V}_i\vec{I}_{ik}^*=P_{ik}+jQ_{ik}$).
    \State 2. Communicate $V_i$, $\theta_i$, $P_{ik}$, $Q_{ik}$ with neighboring buses/agents $k\in\mathbb{N}_i$, and receive the data $V_k$, $\theta_k$, $P_{ki}$, $Q_{ki}$.
    \State 3. Using the collected data, estimate $\hat G_{ik}$ and $\hat B_{ik}$ for each $k\in\mathbb{N}_i$ employing (\ref{lse}).
    \State 4. Compute the required partial derivatives (\ref{partmat1}) using the estimated values $\hat G_{ik}$ and $\hat B_{ik}$, and the latest measurements of $\{V_i,\theta_i,V_k,\theta_k\mid k\in\mathbb{N}_i\}$.
    \State 5. Collect the predicted generation $[\widehat{P}^{G,t+1}_i,\widehat{Q}^{G,t+1}_i]$ and demand profiles $[\widehat{P}^{D,t+1}_i,\widehat{Q}^{D,t+1}_i]$.
    \State 6. Solve the MPC problem (\ref{centmpcchvar}) by \textbf{Algorithm} \ref{alg:admm}. 
    \State 7. From the output of the optimization, obtain the  reactive compensation $u_i^{t+1}$ and implement it for  the next time step $t+1$.
    \EndFor
  \end{algorithmic}
\end{algorithm}

\section{Test Results}\label{5}
\begin{figure}[htbp!]
  \centering
    \includegraphics[width=0.48\textwidth]{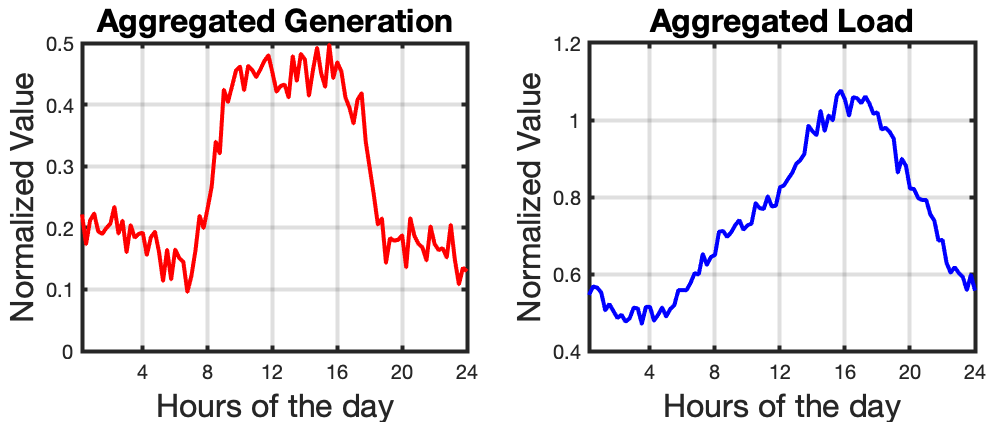}
  \caption{Predicted renewable generation, and  load profile}
  \label{f3}
  \vspace*{-0.2in}
 \end{figure}

We validate our proposed D3PVC framework for (a) Transmission systems: IEEE 30 Bus and IEEE 57 Bus, and (b) Distribution system: IEEE 123 Bus system, by way of tracking a specified reference voltage $V_{\text{ref}}$ under uncertain time-varying loads and renewable generations. 
For our validation purposes, we obtained real-world utility-scale data from \cite{caiso} and plotted the time-varying renewable generation and load profile predictions as shown in Fig.~\ref{f3}. Pypower (or Matpower \cite{zimmermanmatp}) was used to solve the nonlinear ac power flow of the chosen power systems. 
\subsection{Test Case Description: IEEE 30 Bus, 57 Bus and 123 Bus}
\subsubsection{IEEE-30 Bus system} The standard test system with six generators located at buses 1, 2, 13, 22, 23, and 27 is utilized. We include renewable generations (solar and wind) in these generator buses, allowing a mix of conventional and renewable generations, with 50\% coming from renewable sources during their peak generations. Under the representative profiles of Fig.~\ref{f3}, if no control compensation is exercised, the voltage values of multiple buses drop below the recommended $V_{\text{min}} = 0.95$ p.u. as can be seen in Fig.~\ref{f4}(a) (No Control case in red), that has a minimum voltage of around 0.91 p.u.

\subsubsection{IEEE 57 Bus system}
We modified the standard IEEE 57 bus network, which has generators only at buses 3, 8, and 12, by adding generators at buses from 13 to 57. Here again, the generations are taken to be a mix of conventional and renewable ones, with renewable generations serving approximately 50\% of the total load at its peak availability. The predictions of renewable generation and load profiles are same as shown in Fig.~\ref{f3}. Similar to the IEEE 30 bus network, the voltages drop below the desired margin in the absence of any control as depicted in Fig.~\ref{f4}(b) (No Control case in red), thereby necessitating some reactive control compensation for mitigating the voltage drop. 

\subsubsection{IEEE 123 Bus system} IEEE 123 is a large distribution network including overhead, underground line segments, transformers, breakers, capacitor banks, and voltage regulators. In this article, we utilized a balanced version of this test system from \cite{bobo2021second}, and modified it to add renewable generation from buses 27 to 56 and buses 84 to 114. The utilization of balanced distribution system for proof-of-concept is also exemplified in earlier works \cite{magnusson2020distributed,bolognani2014distributed,Baosen,li2019distributed}. The predicted load profiles are same as in Fig.~\ref{f3}; the predicted generation profile is modified to exclude the wind generation. Even then it is observed that voltage values drop below the acceptable range under no control input (see Fig.~\ref{f4}(c) (No Control case in red)).

\subsection{Control Design}
We implemented the proposed D3PVC framework to mitigate the voltage drops observed in all 3 test systems: IEEE 30 Bus, IEEE 57 Bus, and IEEE 123 Bus. Since the algorithm is data-driven and distributed, D3PVC framework collects the measurement data and computes the reactive power required at individual buses (except the slack) (a) to maintain voltage values close to the given reference $V_{\text{ref}} = 1$ p.u., (b) to ensure that the voltage trajectories always remain within the [0.95, 1.05] p.u., and under the reactive power compensation limits of [-0.05, 0.05] p.u. 

\subsubsection{Control under uncertain prediction and modeling errors}
\begin{figure*}[t]
  \centering
    \includegraphics[width=0.90\textwidth]{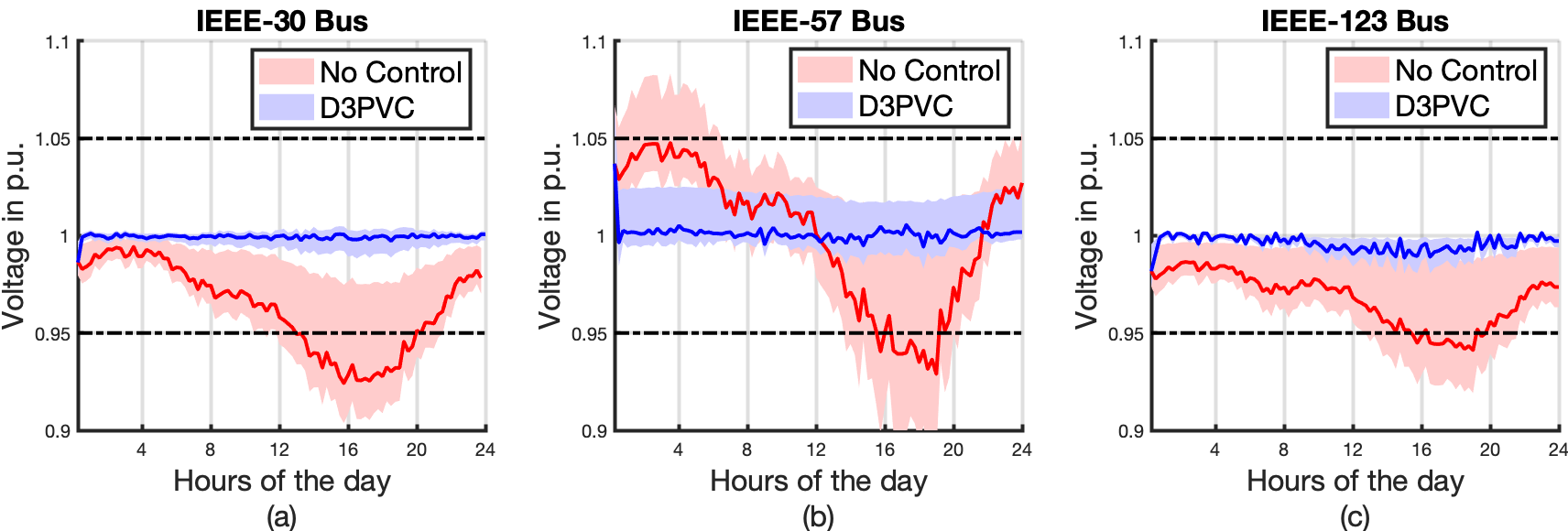}
  \caption{Voltage performance under uncertain prediction and modeling error (a) IEEE 30 Bus, (b) IEEE 57 Bus, and (c) IEEE 123 Bus}
  \label{f4}
 \end{figure*}

\begin{figure}[htbp!]
  \centering
    \includegraphics[width=0.48\textwidth]{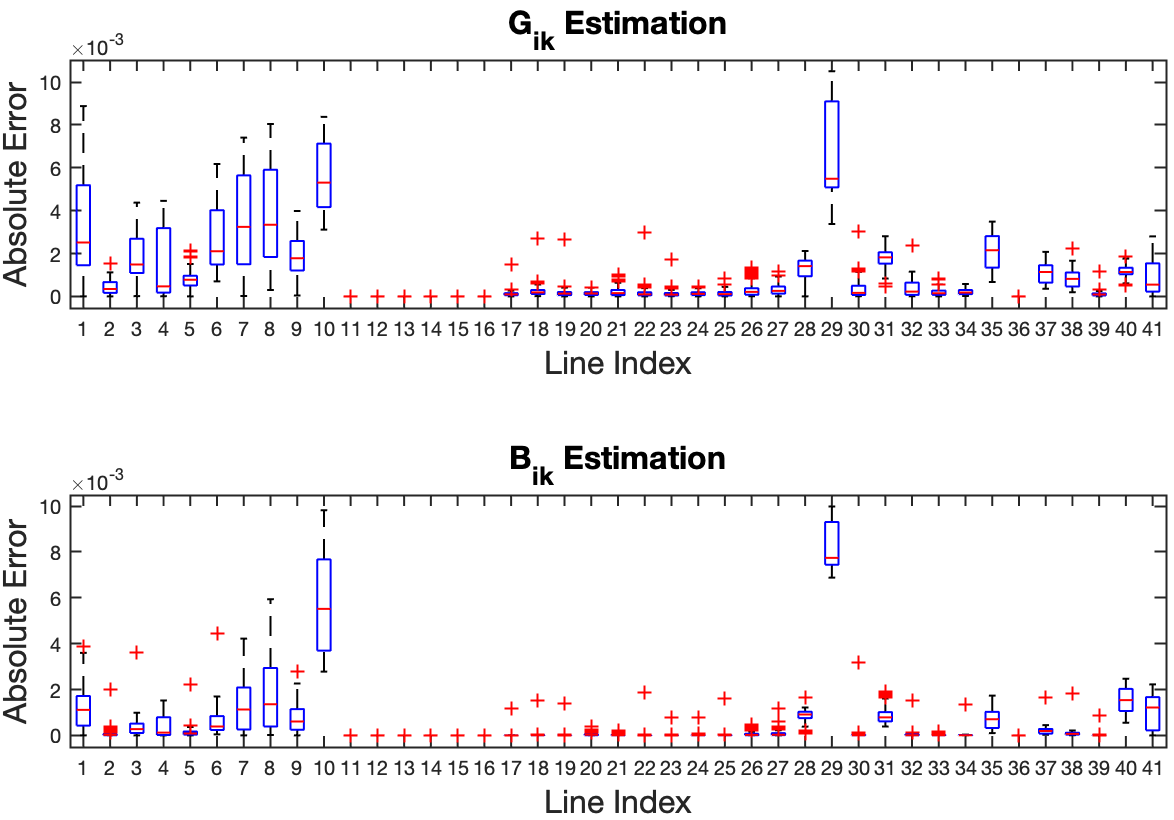}
  \caption{IEEE 30 Bus: Error in estimation of $G_{ik}$ and $B_{ik}$}
  \label{f9}
 \end{figure}

As mentioned earlier, the generation and load profiles are shown in Fig. \ref{f3} are subject to prediction error of 3-5\% in practice (see \cite{caiso}), and therefore we evaluate the performance of the D3PVC scheme under the prediction error of $\pm 5\%$. Moreover, instead of relying on the knowledge of the admittance matrix, we estimated it online which brings in added estimation errors. The estimated errors in $G_{ik}$ and $B_{ik}$ is depicted in Fig. \ref{f9} for IEEE-30 Bus system.
The voltage profiles with D3PVC control is overlayed in blue in Fig. \ref{f4}(a) (IEEE 30 Bus), Fig. \ref{f4}(b) (IEEE 57 Bus), and Fig. \ref{f4}(c) (IEEE 123 Bus). (It should be noted that the variation of voltages over all buses in represented by \textit{red shade} in case of `No Control' and by \textit{blue shade} in case of D3PVC.)
    \begin{figure}[htbp!]
  \centering
    \includegraphics[width=0.48\textwidth]{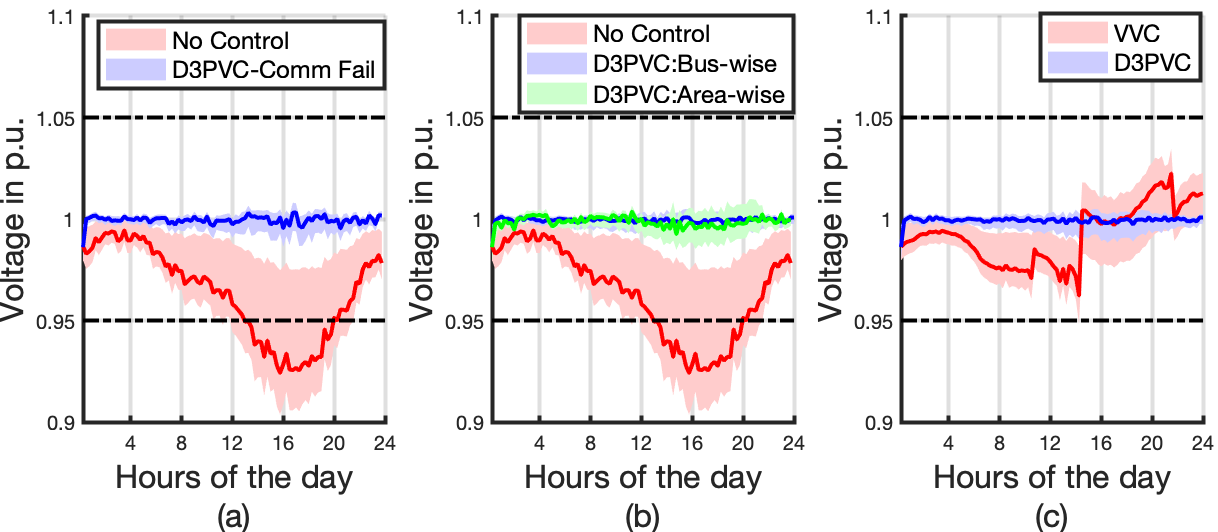}
  \caption{Performance of IEEE-30 Bus (a) Under Communication Failure, (b) Under Area-wise Distributed Control, (c) D3PVC vs. VVC schemes}
  \label{f4a}
  \vspace*{-0.2in}
 \end{figure}
\subsubsection{Control Under failure of communication link}
Here the proposed D3PVC algorithm is tested for random failures of communication links (recall the graph of the communication network is taken to be that of underlying power system). Under dynamic generation, load variation, prediction uncertainty and modeling error we simulated random failures of communication links $l_{ik}\in\mathbb{E}$ between any two neighboring buses $i$ and $k$. In case of such a communication failure, the local controllers at buses $i$ and $k$ eliminate the respective buses from their list of neighboring buses and then proceed with the local control computation. Fig. \ref{f4a}(a) shows the satisfactory voltage performance for the IEEE 30 bus test system (in blue with control vs. in red without control), proving our method's robustness to random failure of communication links. 

\subsection{Sensitivity to Optimization Parameter and Computation time}
\begin{figure}[htbp!]
  \centering
    \includegraphics[width=0.48\textwidth]{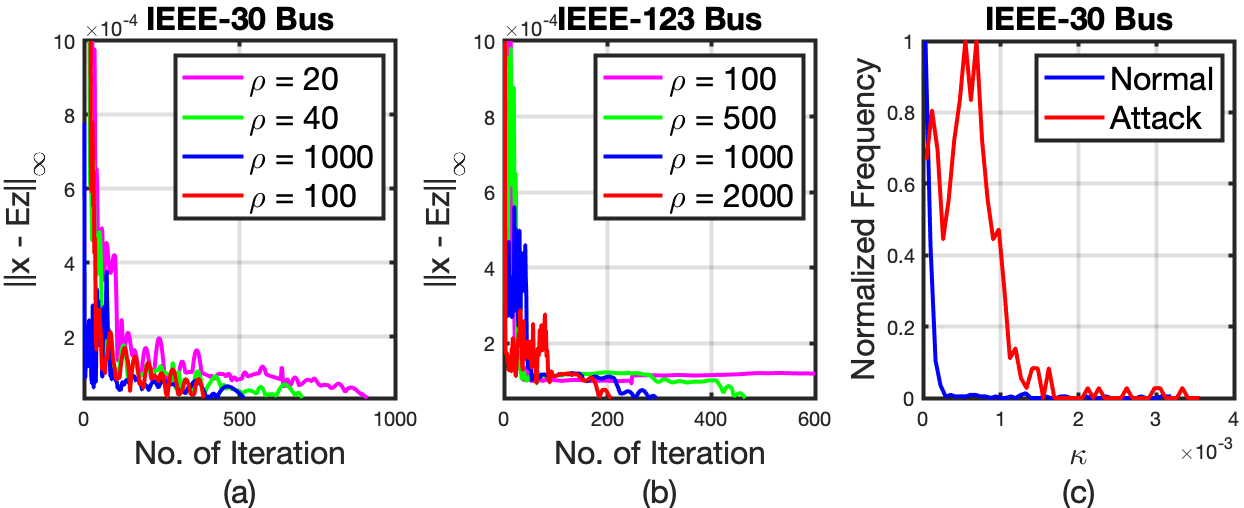}
  \caption{(a) Impact of $\rho$ for IEEE-30 Bus, (b) Impact of $\rho$ for IEEE-123 Bus, and (c) Distribution of $\kappa=(x_{(2,2)} - x_{(6,2)})$ for IEEE-30 Bus}
  \label{fconv}
  \vspace*{-0.2in}
 \end{figure}

We examine the effect of optimization parameter $\rho$ on the convergence speed of Algorithm 2. To balance the speed and accuracy, we set the stopping criteria of ${\lvert\lvert x-E z\rvert\rvert}_{\infty}=\underset{i\in\mathbb{N}}{\textrm{max}}\Big[{\lvert\lvert x_i-E_i z\rvert\rvert}_{\infty}\Big] \leq 3.5\times 10^{-5}$ for IEEE 30 bus test case. The convergence speed is presented under 4 different values of $\rho = 20, 40, 100, 1000$, as shown in Fig. \ref{fconv}.(a). It is found that $\rho=100$ gives the fastest convergence: 380 iterations (approx.). The stopping criteria for IEEE 57 bus test case is set ${\lvert\lvert x-E z\rvert\rvert}_{\infty} \leq 10^{-4}$. In this case we did not observe much variation in convergence speed for $\rho=100$ vs. $\rho=1000$, which took approximately 2300 iterations. In IEEE 123 bus test case, we selected ${\lvert\lvert x-E z\rvert\rvert}_{\infty} \leq 5\times10^{-5}$ as the stopping criteria, and studied the convergence speed for $\rho = 100, 500, 1000, 2000$ (as plotted in Fig. \ref{fconv}(c)). For IEEE 123 bus test case we found that $\rho = 2000$ solves the distributed optimization in least number of iterations.
For our implementation, we run the program in a single machine (intel(R) Core(TM) i7-4790 CPU @ 3.60GHz processor with 16 GB RAM). But, in a real-world implementation, the entire framework can be run parallel with higher computational resources, therefore respective computation time can be divided by number of buses. Given that we observed the average computation time over the computationally expensive time instants to be 124 sec, 1427 sec., and 231 sec. respectively for IEEE 30, 57, and 123 bus test cases, these can come down to the range of 4-25 sec. without communication latency, and 16-100 sec. considering communication latency. 
\subsection{Extension to Area-wise Distributed Control Design}
In Section III, we followed bus-level communication and distributed control computation. But, our D3PVC framework is general enough to also be used for area-wise distributed computation, and as a proof-of-concept, we implemented the area-wise distributed computation for IEEE 30 Bus test case. We divided the IEEE 30 Bus system into three areas connecting through tie-lines with each area containing its own controllers. The distributed control computation is similar to bus-level decomposition, but only the boundary buses get to communicate. The voltage performance is shown in Fig. \ref{f4a}(b), where it is observed that the voltage profile is not very different under bus-wise vs. area-wise distributed computation (blue vs. green).

\subsection{Comparison with Existing Local Volt-Var Control Scheme}
For the sake of completeness, the efficacy of the proposed D3PVC control strategy is compared with the conventional droop-based volt-var control (VVC) scheme. The VVC design is done following \cite{patari2021}, where there is no reactive power injection when bus voltage is in the dead-band range $[0.97,1.03]$ p.u. On the other extreme, if the voltage goes below 0.94 p.u (or beyond 1.06 p.u.). the reactive injection is $+Q_{\text{max}} = 0.05$ (or $-Q_{\text{max}} = 0.05$) p.u. In between, the reactive injection varies linearly from 0 to $+Q_{\text{max}}$ (or 0 to $-Q_{\text{max}}$) in the voltage range $[0.94,0.97]$ (or $[1.03,1.06]$) p.u. The voltage comparison for our proposed D3PVC and VVC schemes are presented in Fig.~\ref{f4a}(c) for IEEE 30 Bus system, which clearly indicates the superiority of D3PVC over VVC (blue vs. red plots).

\subsection{Attack Resiliency of the Framework}
One of the features of a distributed implementation is that it is attack resilient by being not susceptible to single point failures. To demonstrate this, we conducted a simple study and found that the D3PVC framework is promising in rapid anomaly detection using the approach in \cite{anomaly}. In IEEE 30 bus example, bus-2 and bus-6 are neighboring buses, so according to D3PVC protocol, bus-2 has the variable $x_{(2,2)} = \Delta V_2^{t+1}$, while bus-6 has $x_{(6,2)} = \Delta V_2^{t+1}$, meaning the two variables are replicas of the same true variable. Under normal condition, the difference $\kappa = (x_{(2,2)} - x_{(6,2)})$ gradually decreases to 0 as expected, and this difference $\kappa$ follows a distribution as shown in Fig. \ref{fconv}(c) (blue curve). But in case of a random bias injection attack in the communication channel between bus-2 and bus-6, the distribution drastically changes (red curve), which can then be used to alert the operator about possible attacks or unusual behavior. In summary, the proposed D3PVC has in-built redundancy through the incorporation of replica variables that can be leveraged to detect the cyberattacks on communication channels corrupting the data.

\section{Conclusions}\label{6}
This paper developed a novel data-driven distributed predictive voltage control (D3PVC) methodology that allows uncertain and time-varying generations and load demands. The data-driven nature of the framework further implies that a prior knowledge of the line impedance is not required, rather it is estimated online utilizing the data from field measurement devices (voltages and currents). Further, the proposed algorithm only utilizes local measurements and communication, in determining the reactive power injections in the presence of prediction and modeling errors, bus/link failures, and also supports anomaly detection in case of corrupted data through in-built redundancy in form of various replicated variables. The proposed methodology is applicable for both transmission and distribution systems and is not dependent on any assumption on network topology (such as radial or acyclic). The test results applied to IEEE 30-bus, 57-bus transmission systems, and IEEE 123-bus distribution systems validated the performance and robustness of the proposed D3PVC framework against model/generation/demand/network uncertainties/failures/attacks. The algorithm is extendable from bus-level control computation to area-level control computation and can tolerate communication failure. We also demonstrated that D3PVC offers better optimality than standard VVC-based approaches. Future research can explore the extension to unbalanced distribution networks in line with the extensions \cite{patari2022, magnusson2020distributed}, and integration of various robust learning-based methods within the D3PVC framework.

\bibliographystyle{IEEEtran}
\bibliography{IEEEabrv,Bibliography}

\end{document}